\documentstyle[12pt]{article}

\begin{document}

\begin{center}{\bf \Huge Condensation and  Evaporation
 on a Randomly Occupied Square Lattice}
\vskip .5cm Tapati Dutta$^1$,  Nikolai Lebovka$^2$ and S.
Tarafdar$^3$\\ $^1$ Physics Department, St. Xavier's College,\\
Kolkata 700016, India\\ $^2$ Biocolloid Chemistry Institute named
after\\F. D. Ovcharenko\\42, Vernadsky Av., Kyiv, Ukraine\\ $^3$
Condensed Matter Physics Research Centre,\\Physics Department,
Jadavpur University,\\ Kolkata 700032, India.\\Email:
sujata@juphys.ernet.in (S Tarafdar)\\
\end{center}
\vskip .5cm \noindent {\bf Abstract}\\
We study the evolution of an initially random distribution of
particles on a square lattice, under certain rules for `growing'
and `culling' of particles. In one version we allow the particles
to move laterally along the surface (mobile layer) and in the
other version this motion is not allowed (immobile case). In the
former case both analytical and computer simulation results are
presented, while in the latter only simulation is possible. We
introduce growth and culling probabilities appropriate for
condensation and evaporation on a two-dimensional surface, and
compare results with existing models for this problem. Our results
show very interesting behaviour, under certain conditions quite
different from earlier models. We find a possibility of hysteresis
not reported earlier for such models. \noindent

\section{Introduction}
The random percolation problem in 2-d \cite {perc} can be extended
to a number of interesting variations, e.g. the `bootstrapping
model' \cite{boot}, where certain sites are culled leading to a
change in percolation behaviour or the `diagenesis model'
\cite{dia}, with growing as well as culling.
 In the present model we show that an introduction
of simultaneous growing and culling processes, followed by a randomisation
after each time step gives very interesting behaviour.

The model may be applicable to a real situation such as
condensation/evaporation or adsorption/desorption of a layer of
`molecules' at different surface temperatures. We develop two
versions of the model in subsequent sections . In model I, we
assume `growth', that is condensation at sites where there are a
large number of occupied nearest neighbours and `evaporation' at
sites with most neighbouring sites vacant. This algorithm mimics
an attractive interaction between the `molecules'. We find that a
certain coverage ($p_{inv}$) of the surface is invariant,
depending on the exact algorithm employed. An initial
concentration of occupied sites $p>p_{inv}$ leads to complete
coverage, whereas a lower starting point $p<p_{inv}$ leads to zero
coverage, i.e. complete evaporation. In section 2 we discuss the
algorithm for growing and culling and in section 3 a hypothetical
case with a symmetric rule for activated growing and culling is
discussed. In sections 4 and 5, more realistic models II and III
for the condensation/evaporation process as function of vapor
pressure and surface temperature are presented. In the last
section conclusions and future plans are discussed.

\section{Model I}
We start with a square lattice in two-dimensions and sites are randomly
occupied by particles with a probability $p$. Then we grow new particles at
vacancies, or cull particles at occupied sites, according to one of the rules
below.

\vskip .5cm {\bf Rule 1}\\
(a) occupied sites having all four nearest neighbour (nn) sites
vacant and occupied sites with only one nn site occupied are
culled. (b) A new particle grows, at a vacant site with four nn
positions occupied, and also at a vacant site where only three nn
sites are occupied.

\vskip .5cm {\bf Rule 2}\\
(a) Only particles with four vacant nn are culled. (b) Only sites
with four  occupied nn `grow' a new particle.

\vskip .5cm {\bf Rule 3}\\
(a)Only particles with four vacant nn are culled. (b) A new
particle grows, at a vacant site with four nn positions occupied,
and also at a vacant site where only three nn sites are occupied.

\vskip .5cm {\bf Rule 4}\\
(a) Occupied sites having all four n.n. sites vacant
and occupied sites with only one nn site occupied are culled.
(b) Only sites with four nn occupied `grow' a new particle.

\vskip .5 cm

In all the above rules particles or vacant sites with 2 nn
occupied are left undisturbed. Figure 1 illustrates the rules pictorially.

After completing the grow-cull operations, the concentration of
particles changes from an initial value $p_i$ to a final value
$p_f$. We now randomise the positions of the remaining  sites over
the whole lattice, and repeat the grow-cull operations with $p_f$
becoming the new $p_i$. The results of the above procedure for the
four different rules outlined above can be determined
analytically.

For each of the four rules we find an initial coverage $p_{inv}$
which remains invariant after repeating the steps of
growing-culling followed by randomisation.  $p_{inv}$ corresponds
to the coverage for which the probability of growth equals the
probability of culling for the particular rule employed. For some
other starting concentration say $p_0>p_{inv}$, we would get a new
$p=p_1$ which is $>p_0$, since the growth probability exceeds
culling probability. The randomisation that follows makes $p_1$
the new initial coverage, which after growing-culling gives a
still larger $p_2$, and so on. So the coverage approaches 1.0 or
100 \%, as shown in figure 2.

For $p_0<p_{inv}$, on the other hand $p_1 < p_0$ and the system
evolves towards $p=0$, i.e. zero coverage, $p_{inv}$ is an
unstable fixed point for the system, whereas $p=0$ and $p=1$ are
stable fixed points.

The processes described above can be studied by computer
simulation  as well as analytically. Using a parallel algorithm
for growth and culling gives a result exactly in agreement with
the calculated result, while a sequential algorithm gives
different results as expected.

The stable fixed point can be determined as follows. For a
coverage $p$, the probability of growing and culling $P_{gr}$ and
$P_{cl}$ for the four different rules can be written as follows.
For Rule 1
\begin{equation} P_{cl} = p (1-p)^4 +4p^2 (1-p)^3 \label{e1} \end{equation}
\begin{equation} P_{gr} = p^4 (1-p) + 4p^3 (1-p)^2 \label{e2}\end{equation}
In eq.1 the 1st term on the right is the probability  for an
occupied site having 4 vacant nn, in  a random distribution. since
$p$ is the probability of a site being occupied, and $(1-p)$ is
the probability of being vacant. The other terms can be written
down similarly with proper weight factors.

For the other rules, we have analogous relations. For rule 2
\begin{equation} P_{cl} = p (1-p)^4 \label{e3} \end{equation}
\begin{equation} P_{gr} = p^4 (1-p) \label{e4}\end{equation}
For Rule 3
\begin{equation} P_{cl} = p (1-p)^4 \label{e5}\end{equation}
\begin{equation} P_{gr} = p^4 (1-p) + 4p^3 (1-p)^2 \label{e6}\end{equation}
and Rule 4
\begin{equation} P_{cl} = p (1-p)^4 +4p^2 (1-p)^3 \label{e7}\end{equation}
\begin{equation} P_{gr} = p^4 (1-p) \label{e8}\end{equation}

$p_{inv}$ is easily obtained for any of the rules by setting
$P_{cl}=P_{gr}$ and solving for $p$.

For Rules 1 and 2, which are symmetric, we find $p_{inv}=0.5$,
while for Rule 3 and Rule 4 $p_{inv}$ has the complementary values
0.32 and 0.68 respectively. The same values are obtained for a
steady coverage by carrying out a computer simulation of the
processes described by the rules.

\section{Activated growth and culling}
From physical considerations it is expected that - the probability
of condensation on a surface decreases with increasing surface
temperature, and probability of evaporation increases \cite{jay}.
If we assume that rule 3 holds at high temperatures and rule 4 at
low temperature, we may introduce a hypothetical temperature
dependent growth and culling rule as follows.

\begin{equation}
P_{cl} = p (1-p)^4 + 4 exp(-E_0/kT) p^2 (1-p)^3 \label{e9}
\end{equation}
\begin{equation}
P_{gr} = p^4 (1-p) + 4 (1 - exp(-E_0/kT)) p^3 (1-p)^2 \label{e10}
\end{equation}

The temperature dependence enters through the exponential factor
with $E_0$ as an energy characteristic of the system.

The new rule can be seen to reduce to Rule 3  for  $T \rightarrow
\infty$, and to Rule 4 for $T \rightarrow 0$. The fixed point
$p_{inv}$ goes accordingly from 0.32 to 0.68 as temperature is
lowered from a very high value compared to a characteristic
temperature $T_0=E_0/k$. $p_{inv}$ can be determined by equating
the growth and culling probabilities as before at different
temperatures.

The exponential temperature dependence introduced here for
evaporation is realistic, but the rule for condensation is purely
hypothetical, introduced to make the equation symmetric. In the
next section we discuss a more realistic approach. The final coverage in this case 
is either $0$ or $1$ depending on the temperature and the initial coverage. 

\section{ Condensation and Evaporation }

In this section we study the condensation/evaporation  problem
using the approach developed in the preceding section. We find
that computer simulation of the problem is instructive, and may
lead to a reinterpretation of some existing ideas.

The process of condensation/evaporation or adsorption/desorption
is described traditionally by two different sets of models -- one
for a mobile adsorption layer and one for an immobile
layer\cite{book1,book2}. Probabilities for sticking and
evaporation on a two-dimensional monolayer are specified according
to the physics behind the model.These are functions of the
temperature, superincumbent pressure and the existing coverage. At
equilibrium, the sticking and evaporation probabilities are set
equal and the resulting equation is solved to get the equilibrium
coverage at that temperature and pressure.

The simplest mobile layer model is a two-dimensional ideal gas,
and the improved versions include interaction between particles,
similar to a two-dimensional Van der Waal gas. The so-called
`immobile layer' models introduce a sticking probability, depending
on how long a molecule in the vapor above the surface is in
contact with a surface site. The simplest `immobile model' is the
 Langmuir equation derived as follows.

For vapor condensation the particle flux, i.e. the number of
particles deposited per unit time per unit surface area  is equal
to:
\begin{equation}
c=\frac{P\lambda}{h} \label{e13}
\end{equation}
where
 $P$ is the pressure, $$\lambda=\sqrt{h^{2}/(2\pi mkT)}$$ is de
Broglie length, $h$ is Planck constant.

Evaporation probability from saturated surface (at $p=1$) may be
approximated by \cite{jay} :
\begin{equation}
d=\frac{kT}{h\lambda^{2}}\exp(-E_{e}/kT) \label{e14}
\end{equation}

Condensation probability is set equal to evaporation probability,
giving equilibrium:
\begin{equation}
\frac{P\lambda}{h}(1-p)=\frac{kT}{h\lambda^{2}}\exp(-E_{e}/kT)p
\label{e15}
\end{equation}
or
\begin{equation}
\frac{P\lambda^{3}}{kT}(1-p)=\exp(-E_{e}/kT)p \label{e16}
\end{equation}
and we have the simple Langmuir equation
\begin{equation}
Pb=\frac{p}{1-p} \label{e17}
\end{equation}
where
$$b=\frac{\lambda^{3}}{kT}\exp(E_{e}/kT),$$
$E_e$ is activation energy for evaporation.

In this approximation it is assumed that evaporation energy is the
same for  all configurations.

A modification of this model is the Fowler-Guggenheim model
\cite{book2, jay}, one form of this is given below. Here the
evaporation probability for a particle depends on the number of
occupied neighbors. Each of the neighbors exerts an attractive
force on the particle, which must be overcome for evaporation. We
introduce the following parameters for convenience, -  $f$ is a
pressure parameter given by$$f = P \Lambda^3/gkT $$ $$\Lambda =
\sqrt{T^\star}\lambda$$ and $$g =exp(-E_e/kT)= exp(-T_0/T)=
exp(-1/T^\star) $$ where $T_0=E_e/k$ is a characteristic
evaporation temperature, $T^\star=T/T_0$ is a reduced temperature.

The condition for
condensation rate to equal evaporation rate is now
\begin{equation}
f(1-p)/(T^{\star})^{5/2} = p^5g^4 + 4p^4g^3(1-p) + 6p^3g^2(1-p)^2 + 4p^2g
(1-p)^3 + p(1-p)^4 \label{e18}
\end{equation}
The successive terms on the RHS of equation (\ref{e18}) represent
probabilities for a particle to have 4, 3, 2, 1 and 0 nearest
neighbours in a random distribution On rearrangement this reduces
to
\begin{equation}f(1-p)/(T^{\star})^{5/2} = gp(1+pg-p)^4 \label{e19}\end{equation}

\section{Condensation/evaporation from a different viewpoint:
Models II and III}

In the above models the question of mobility of the adsorption
layer is not introduced explicitly, though the FG model is
classified as an immobile model. Moreover the probabilities for
evaporation considered in eq.(\ref{e18}). are valid only as long
as the distribution is random. For a strictly immobile layer the
distribution ceases to be random once the site-dependent
evaporation starts. We formulate the problem so that the lateral
mobility if present is introduced explicitly, and we can look at
both mobile and immobile situations within the same framework. As
in the previous section, we start with a two-dimensional lattice
with a certain fraction occupied randomly by the particles. The
adsorbate particles are also present as vapor above the surface
and the pressure and temperature have a key role to play.

We visualise the condensation/evaporation (or
adsorption/desorption) as a two-step process, there is one
characteristic time for the sticking and evaporation and another
for lateral diffusion of the molecules over the surface. For the
`mobile interface' situation, the timescale for lateral motion is
very small, and each condensation-evaporation step is followed by
a complete randomisation. The other extreme is the `immobile
interface', here we drop the randomisation  process altogether. It
is also possible to consider intermediate situations where the two
characteristic times are comparable.

According to this picture, setting the sticking probability equal
to the evaporation probability at equilibrium, is valid for the
mobile situation only, not the immobile case. This is because, if
the surface molecules cannot move laterally, after one round of
growth and evaporation, the expressions for
condensation/evaporation probability are no longer valid, because
the distribution is no longer random.  So for the mobile case, we
have a result similar to the Fowler-Guggenheim formalism, but with
a different interpretation. Our solution for the final coverage
depends on the initial coverage, besides temperature and pressure.
The coverage isotherm of the system may follow a different path
during increasing and decreasing pressure showing hysteresis. This
mobile case, can be worked out by analytical calculation as well
as computer simulation.

The `immobile surface' case, cannot however, be calculated
analytically, as we do not know the condensation/evaporation
probabilities after one round of growth and culling, since the
system has lost its random distribution. But we can still simulate
the system on a computer. We have two cases - firstly the
Fowler-Guggenheim equation in the form of eq.(16), we call this
model II. Secondly  we take a situation where  only particles with
one occupied neighbor or none at all (isolated particles) are
allowed to evaporate (as in rule 1a, section 2), we call this
model III. The corresponding equation is
\begin{equation}
f(1-p)/(T^{\star})^{5/2} = 4p^2 (1-p)^3 + p(1-p)^4 \label{e20}
\end{equation}
Temperature and pressure dependence of sticking probabilities are
assigned as in FG model eq. (\ref{e18}).

\subsection{Mobile interface layer -- analytical study}
Here, there is a complete randomisation  after each
growing-culling sequence, so it is meaningful to equate the
probability of growth to the probability of culling and solve for
the invariant coverage. We can look at the process as an iteration
of the following two steps
\begin{equation}
p_f = p_i + P_{gr} -P_{cl} \label{e21}
\end{equation}
and
\begin{equation}
p_i = p_f \label{e22}
\end{equation}
here $p_i$ is the initial coverage and $p_f$ the final. Also
$P_{gr} = P\lambda (1-p)/h$ and
$P_{cl}=wT^{\star^{2}}exp(-1/T^\star)$. Here $w$ is a parameter,
with a suitably chosen arbitrary value. This iteration done numerically
gives the same result as solving analytically
\begin{equation}
P_{gr} = P_{cl} \label{e23}
\end{equation}
and  also agrees with an explicit  computer simulation of the
process. At certain values of temperature and pressure, there are
three solutions for the coverage. The middle one is an unstable
fixed point(UFP), and the other two stable fixed points. So a
starting coverage above the upper fixed point leads to the
coverage stabilising at the upper fixed point, whereas if we start
with a coverage below UFP we end up at the lower fixed point. The
computer simulation shows exactly the same behaviour. Figures [3a]
and [4a] show the results for eqs (\ref{e19}) and (\ref{e20}). The
difference with the standard thermodynamic treatment is as
follows. If we start from a very low coverage say at a temperature
$0.5$ and pressure $10^{-5}$ and gradually increase the pressure,
according to earlier FG model the coverage increases as shown in
fig3a until a phase transition to the upper fixed point takes
place according to the equal area Maxwell's rule
\cite{MaxwellRule}. In the present model however, stable coverage
depends on initial $p$ as well as $T^\star$ and $f$. If we start
with a very low $p$ at low pressure, the system follows the same
path as FG initially but undergoes the phase transition later as
shown in fig.3a (pt. B) at a pressure where the lower fixed point
meets the UFP. It then follows path BC. While decreasing the
pressure from a higher value the system follows the different path
shown in the figure, with a phase transition where the upper fixed
point coincides with UFP (pt.D). So here we have a marked
hysteresis. Hysteresis in adsorption/desorption is usually
attributed to presence of pores \cite{book1}, but we see here that
it may also have a different origin. In model III (see fig.4a) the
reverse path starting from high  pressure and complete coverage
shows no decrease in $p$ but continues to low pressures with
$p=1$. This is because in this model there is no evaporation
unless some sites are vacant. Model II looks more realistic under
normal conditions. It must be noted that this is quite different
from the usual form of Fowler-Guggenheim, where one would not
expect hysteresis, and the present version considers a {\it
mobile} layer.

\subsection{Computer simulation of the mobile case}

In the mobile case, the adsorbed molecules can move laterally on
the surface. In model II, the physical situation simulated is
described by eq. (16). A two-dimensional square lattice of unit
spacing and size $300 \times 300$, is occupied randomly with an
initial coverage $p_{initial}$. Every occupied site is assigned
the value $1$, empty sites are assigned the value $0$. The
occupied sites are then culled paralally with a probability
determined by the number of their occupied nearest neighbours. The
site having n occupied neighbours, has the culling probability
$p^{n+1}g^{n}$, where $p$ is the occupation probability and
$$g=exp(-1/T^{*})$$ The vacant sites are  filled with a
probability $f/(T^{*})^{5/2}$ where $f$ is the pressure parameter
defined earlier. After one round of growth and culling is
complete, the concentration of the occupied sites $p_{final}$ is
calculated.

In the next time-step, the $p_{final}$ of the previous time-step
becomes the new $p_{initial}$. The square lattice is then randomly
occupied afresh with this $p_{initial}$. A complete time-step
begins with the random occupation of all sites with a
$p_{initial}$ and ends with the assignment of the $p_{final}$ to
the $p_{initial}$ of the next time-step. This iterative process
stops when $p_{final}=0$ or $1$  or $p_{final}$ saturates with
increasing time to a definite value. In the simulation, we checked
upto 50,000 time-steps. The `mobility' of the molecules is
simulated by the randomization of the concentration $p_{initial}$
in the beginning of every time-step.

In model III described by eq.(\ref{e20}), the same iterative
process described in model II is carried out, except for the
condition of culling. In this case an occupied site is culled with
the probability $g$ if the sum of its nearest neighbours is less
than two.

All the sites in both the models are updated parallelly and
periodic boundary conditions are applied both along the x- and y-
directions. Both the models are studied over an effective
temperature range of $0.5$ to $3$ and with $f$ varying from
$10^{-4}$ to $10^{2}$, for the entire range of $p_{initial}$ from
$0$ to $1$. The simulation results agree to within $10^{-3}$ of
the numerical results. These are presented in the figures
($3a,3b,4a,4b$ and $5$).

A significant difference between models II and III is evident from
figs.3a and 4a, showing isotherms for different temperatures. In
model II there is a critical temperature above which there is no
phase transition, but in model III there is always a phase
transition. Fig.5 shows how the coverage evolves with time for
$T^\star = 0.5$ and $f = 0.02$ for $p_{initial}$ varying from 0 to
1.

\subsection{Immobile interface layer}

Let us  now make the adsorbed molecules immobile. In this case
obviously, analytical calculation is not feasible. If we start
with an initial random configuration with a certain coverage, as
soon growth and culling at preferential locations starts we can no
longer calculate probabilities for further evolution exactly. So
here we resort to computer simulation. The simulation of the
`immobile' case of the models II and III follow the same iterative
process, except for the randomization of the sites with the given
probability at the beginning of every time-step. Here a new round
of growth and culling commences on the geometrical configuration
reached at the end of the last time-step. Since in the preceding
section we saw that computer simulation results agree with
calculations, we are confident that the simulation gives reliable
results, with the system size  and algorithm used.

We find that the results are quite different from the $mobile$
case. Whereas in the mobile case, we found definite fixed points
where the coverage converged regardless of the exact starting $p$
(see fig.5), here for certain ranges of the temperature and
pressure parameters, we get a different stable coverage for even
closely spaced initial $p$ values. Figs.(6a \& 6b)  show the time
evolution of the coverage for the immobile system for typical
situations under quite different conditions of pressure. In
fig.6a. for $p_{initial}$ above 0.3 the coverage always goes to 1
at rates depending on the initial coverage, but for lower starting
points it stabilises to different values for  each $p_{initial}$.
In a similar study of the $mobile$ case one always ends up at
either the upper or the lower fixed point (fig.5). But fig.6a
shows that for the immobile case, initially culling dominates, for
large $p_{initial}$ there is a rapid drop in coverage within a few
hundred time-steps. After that sticking catches up, and for $p
>0.3$ (approximately) there is a steady increase, and $p
\rightarrow 1$ linearly with a slope decreasing as $p_{initial}$
decreases. For $p_{initial}$ lower than $\sim 0.3$ $p$ apparently
saturates, showing no variation upto 50,000 time-steps. But when
the same situation is studied at a much lower $f$ ($10^{-10}$),
(fig.6b), $p_{final}$ always saturates to different values
depending on $p_{initial}$, over the time-scales checked. For
example, in fig.6a., for $p_{initial}=0.9$, $p_{final}=1$ whereas
the same $p_{initial}$ in fig.6b. saturates to $p_{final}$ of
0.8979. This suggests that even for the same $p_{initial}$ there
must be a definite combination of pressure and temperature where
there is a change in the behaviour -- $p_{final}$ reaching the
stable fixed point $1$ or a constant value $<1$. Further
investigations probing the exact phase diagram is
in order and will be done in the near future.

All the above figures are the results on model III. Similar studies on 
model II show no significant qualitative difference with model III 
except for the exact numerical values.

\section{Conclusions}

We have investigated the behaviour of a two-dimensional
surface  layer explicitly allowing or forbidding lateral motion of
the particles. We find that the cases for a mobile and an immobile
layer show quite different characteristics, moreover none agrees
fully with the Langmuir or FG model. The fixed points for our
model II with `mobile' particles, are the same as the FG
solutions, which are supposed to be valid for an {\it immobile
layer}. Details of the evolution of the surface coverage from low
to high pressure are however different. For the immobile layer the
results are again different. These results are similar to what was
reported in a previous work attempting to simulate diagenetic
processes of restructuring in sandstones \cite{dia}. However the
two-dimensional character is more appropriate in case of the
condensation/evaporation problem. It will be interesting if
experimental support for the behaviour described here can be
demonstrated. Hysteresis has been observed for adsorption on
porous surfaces \cite{book1}, but here we see that competition
between site-specific sticking and evaporation may also cause
hysteresis.

We can simulate  intermediate behaviour between mobile and
immobile layers, by allowing the particles to execute a random
walk for some time on the surface before the next grow-cull
operation. Making the time for the walk extremely large will
correspond to the complete randomisation done here. We have plans
to study this in future.

Recent work on  adsorption/desorption problems [8-13] show that considerable theoretical
and experimental studies are being done on such problems. They are
of interest in modern devices which use surface properties
extensively and also environment relate problems such as
explaining the ozone hole in polar regions \cite{book1,hay}.
Complicated mathematical and computer simulation methods such as
density functional theory and molecular dynamics simulations are
being used. In this work we show that very simple monte carlo
simulations also reveal some interesting features and may turn out
to be quite useful in shedding some light on such problems. In
conclusion the growth-culling-randomisation extension of the
standard 2-d percolation problem promises to yield more new and
interesting results.

\section{Acknowledgement}
DST, Govt. of India and   Ministry of Ukraine for Education and
Science,  are gratefully acknowledged for grant of an
Indo-Ukrainian collaboration project. Authors thank S S Manna and
J K Bhattacharjee for suggestions and discussion on this subject.
Authors are grateful to S N Bose National Centre for Basic
Sciences for extending their computer facilities.

\newpage
\noindent {\bf figure Captions :}

\vskip .5 cm

Figure 1. The growth and culling processes are illustrated, the
rules are implemented as follows : rule 1 - (a), (b) (d) and (e),
rule 2 - (a) and (e), rule 3 - (a), (d) and (e), rule - 4 (a), (b)
and (e).

\vskip .5 cm

Figure 2. This is a schematic diagram showing how the system
evolves towards the fixed points, here the unstable fixed point is
p = 0.5, valid for rules 1 and 2.

\vskip .5 cm

Figure 3a. Coverage vs. pressure for the mobile case (model II)
with $T^*=0.5$ If we start with a very low coverage ($p$) and
gradually increase pressure, $p$ increases along the curve as
shown upto B. After this it undergoes a phase transition to $p=1$
along BC. On decreasing pressure, it follows the path CDA showing
a well marked hysterisis.

\vskip .5 cm

Figure 3b. Plot of final coverage vs. pressure for different
$T^*$. \vskip .5 cm

Figure 4a. Plot of coverage vs. pressure for $T^{*}=0.5$ for model
III. Here the final coverage depends on initial coverage. For
example, if we start at a point with coverage $<B$ the final
coverage reaches point B. For any starting coverage between points
B and C, the final coverage is B. For starting coverage above C,
the final coverage is point D ($p=1$). Starting with a low
coverage, if pressure is gradually increased, the coverage
proceeds along the curve OBE where it undergoes phase transition
to F.

\vskip .5 cm

Figure 4b. Plot of final coverage vs. pressure for different
$T^{*}$ (model III).

\vskip .5 cm

Figure 5. Plot of coverage vs. time for the mobile case. For
initial coverages $>0.68$ (dotted lines) the final coverage is
$p=1$. For coverages $<0.68$ (solid lines) the final $p$ is the
lower fixed point.

\vskip .5 cm

Figure 6a. Plot of coverage vs. time for the immobile case in
model III for$T^{*}=4$ and $f=10^{-4}$. For initial coverage
$<0.35$, the final coverage saturates to different values. But for
larger initial coverage, the system approaches $p=1$. The numbers
on the curves indicate the initial coverage. The y- axis is 
shifted from $t=0$ to show clearly the initial drop of the coverage
at early times.

\vskip .5 cm

Figure 6b. Plot of coverage vs. time for the immobile case in
model III for $T^{*}=0.3$ and $f=10^{-10}$. Here the final coverage
saturates to different values for all the initial coverages
studied over 50,000 time-steps.


\begin{thebibliography}{99}
\bibitem{perc} D Stauffer and A Aharony,
{\it Introduction to Percolation Theory}
 (Taylor and Francis, London, 1994)
\bibitem{boot} S S Manna, Physica A {\bf 261} (1998), 351.
\bibitem{dia} S S Manna, T Datta, R Karmakar, S Tarafdar, Int. J. Mod.
Phys. C, {\bf 13}, (2002),319.
\bibitem{book1} {\it Surface Chemistry}, E M McCash ({\it Oxford University Press, New York 2001})
\bibitem{book2} {\it On Physical Adsorption}, S Ross and J P Olivier ({\it Interscience, John Wiley and
Sons, New York 1964})
\bibitem{jay}  {\it Chemistry of interfaces}, M J Jaycock and G D  Parfitt,
({\it Wiley, Chichester, 1981}).
\bibitem{MaxwellRule} {\it Statistical Mechanics}, K Huang  ({\it Wiley, New York, 1963}).
\bibitem{gupta} P Gupta, P A Coon, B G Koehler, S M George, J. Chem, Phys.{\bf 93}, (1990), 2827.
\bibitem{hay} D R Haynes. N J Tro and S M George, J. Phys. chem. {\bf 96}
(1992), 8502.
\bibitem{toth} J Toth, Advances in Colloid and Interface Science,
{\bf 55} (1995), 1.
\bibitem{sing} K Sing, Colloids and Surfaces {\bf A187-188} (2001), 3.
\bibitem{dabrowski} A Dabrowski, Advances in Colloid and Interface Science
{\bf 93} (2001), 135.
\bibitem{steele} W Steele, Applied Surface Science {\bf 7820} (2002), 1.

\end{thebibliography}
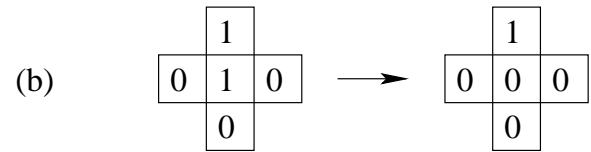
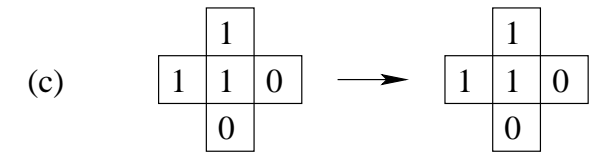
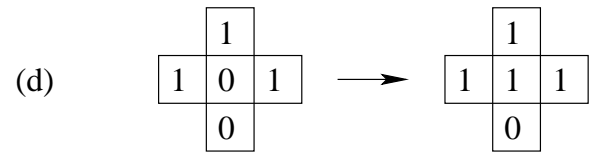
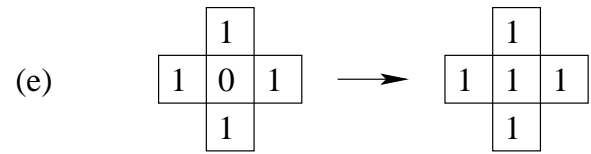
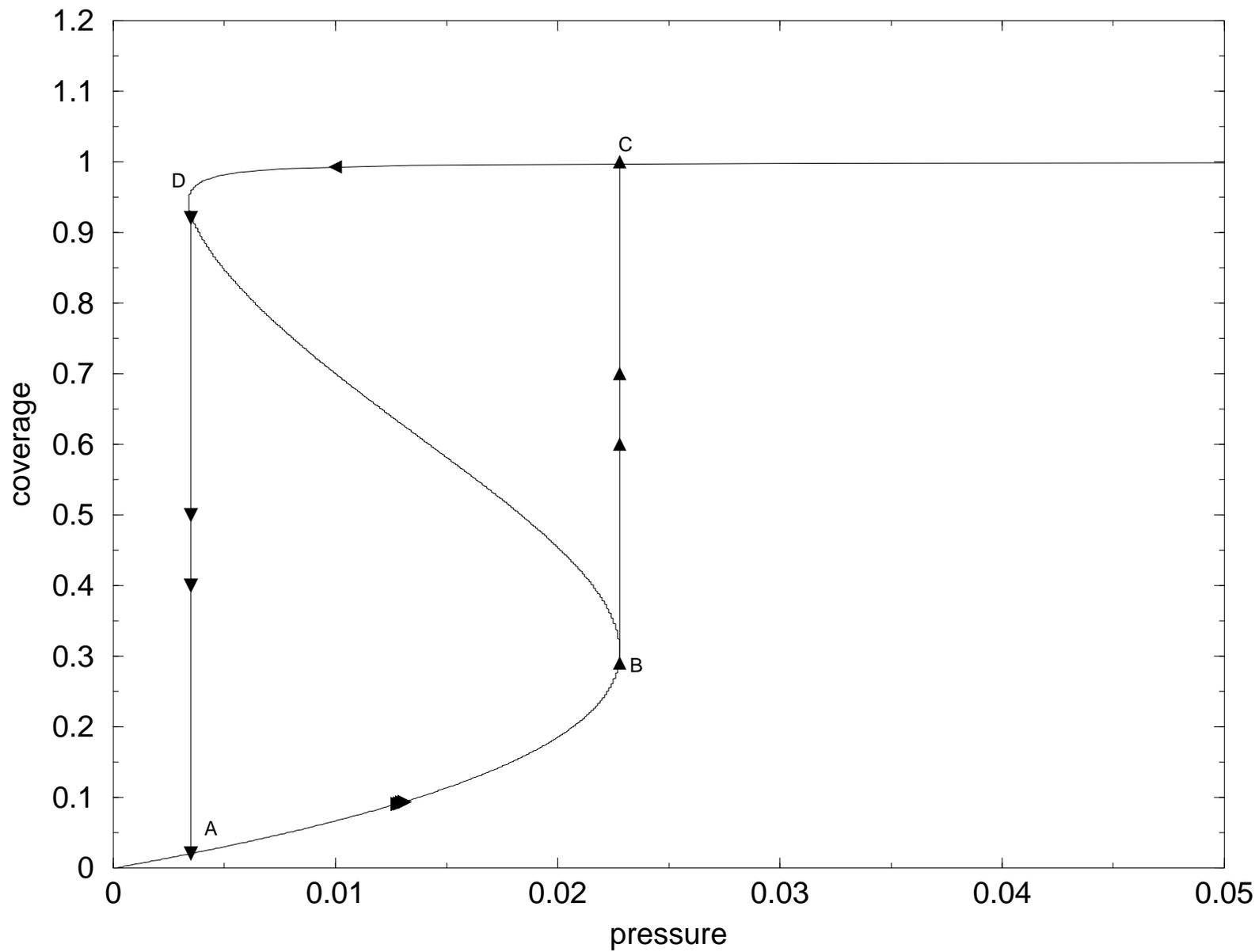
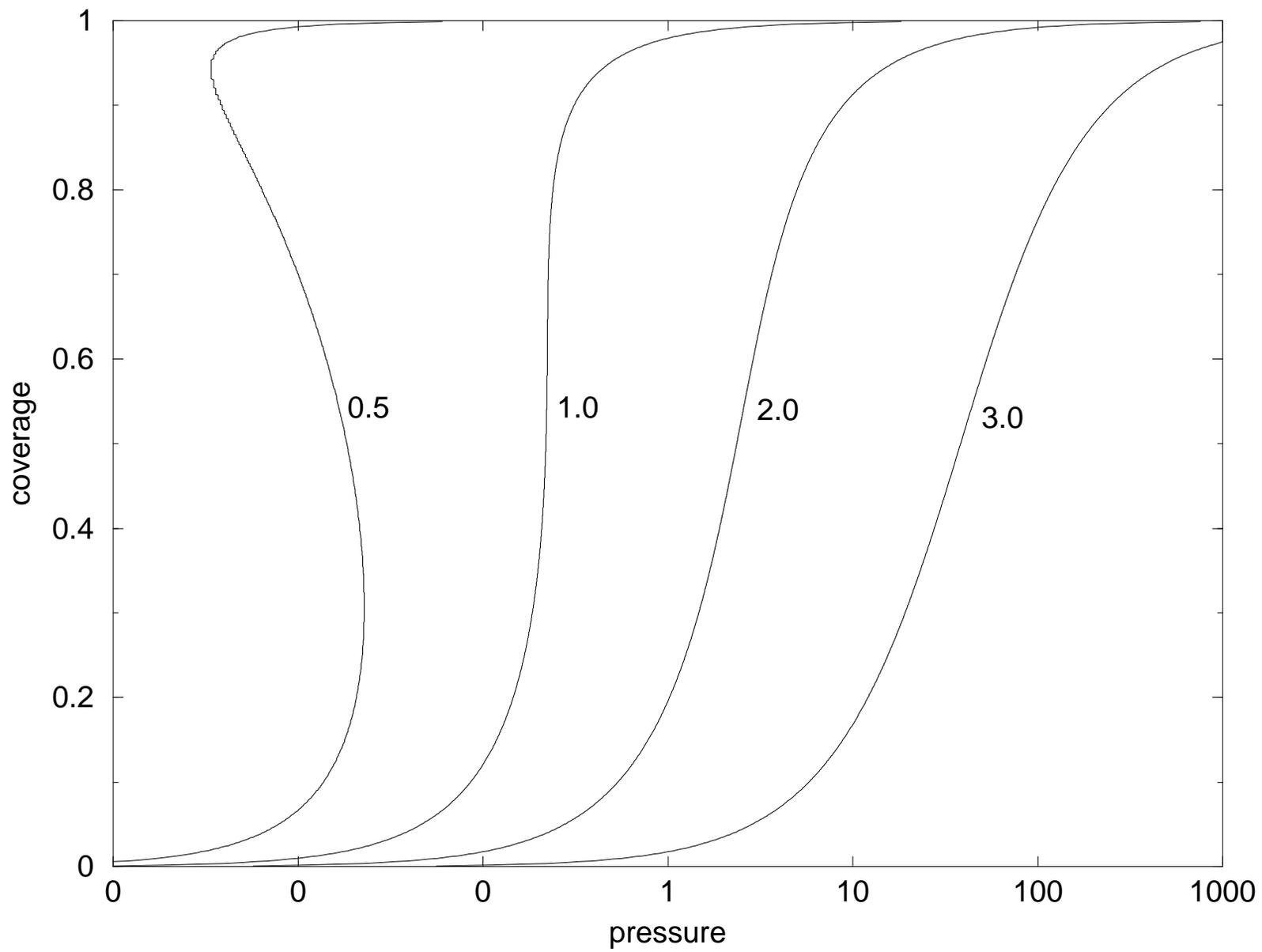
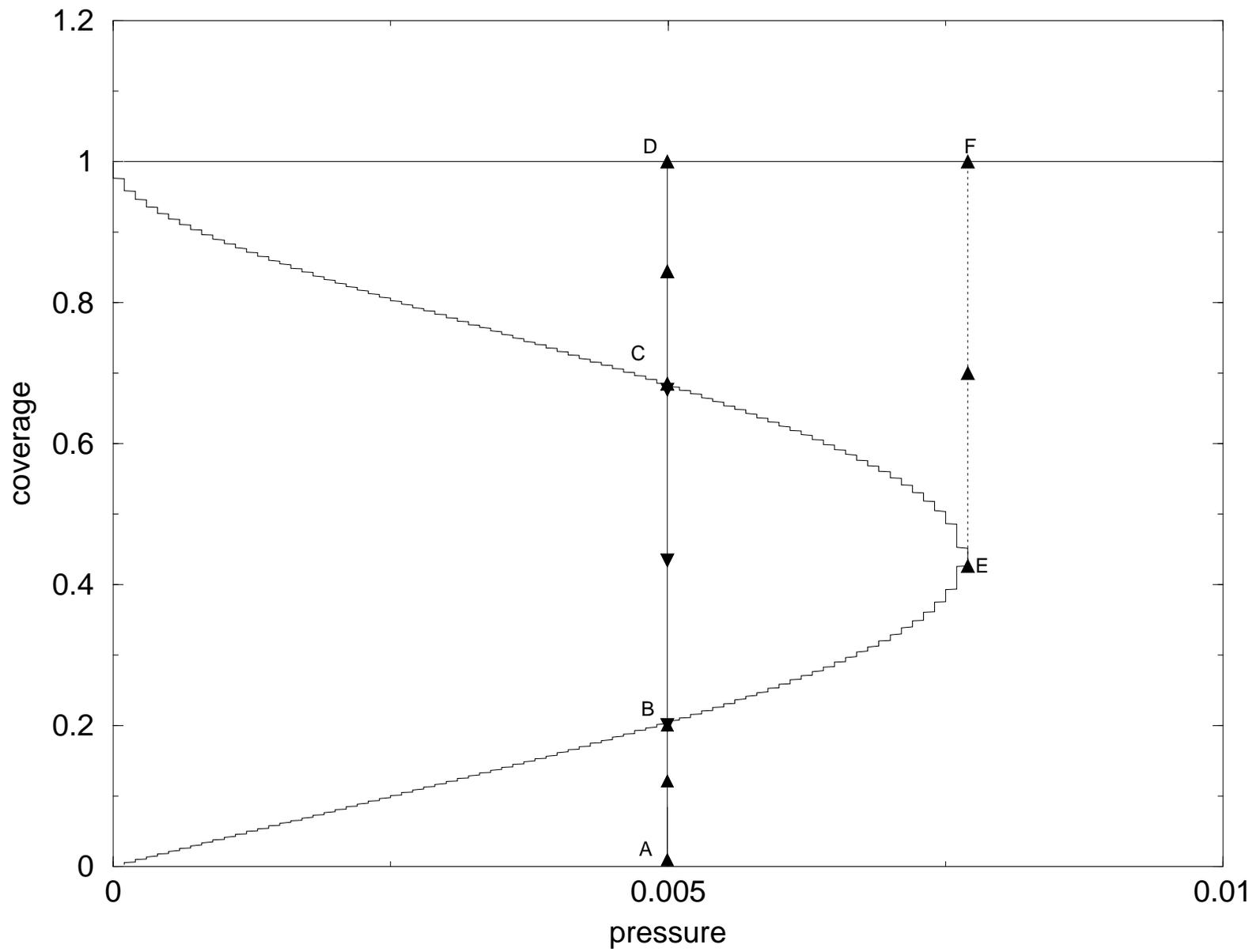
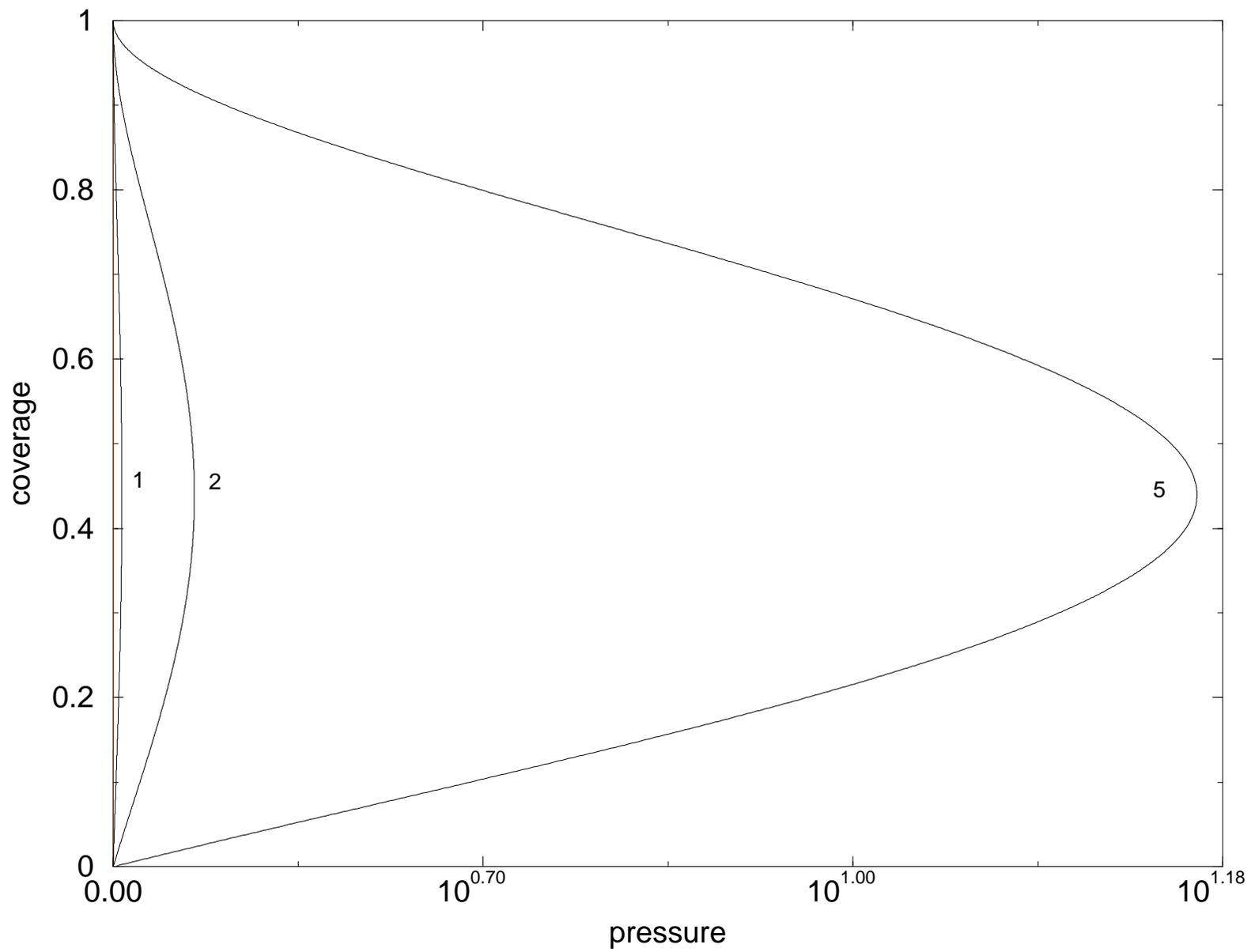
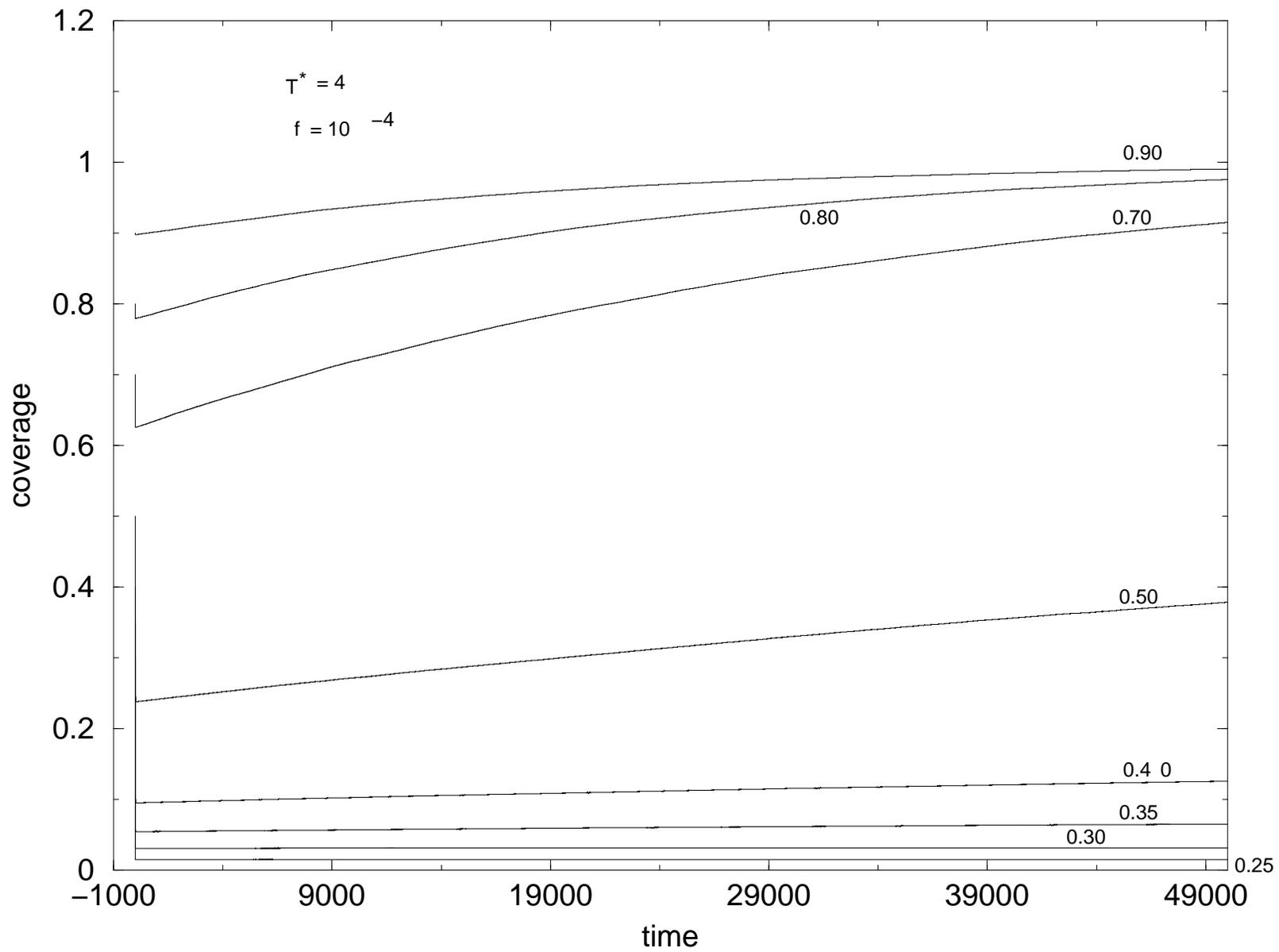
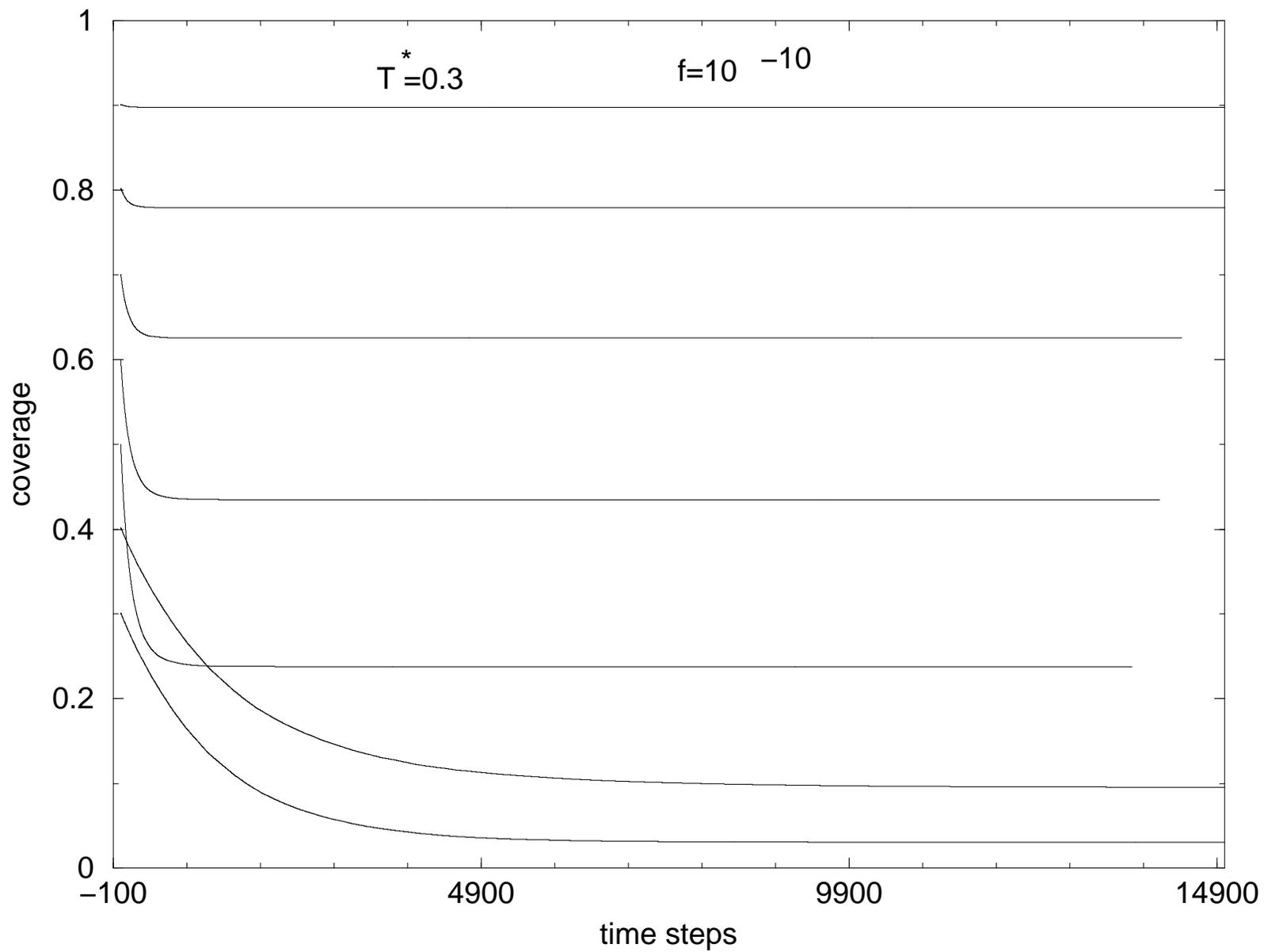
\end{document}